\documentstyle[twoside,fleqn,espcrc2,epsfig]{article}

\title{A fixed-point action for the lattice Schwinger 
model\thanks{Supported by Fonds zur F\"orderung der Wissenschaft\-
li\-ch\-en Forschung in \"Osterreich, Project P11502-PHY.}}

\author{C.~B. Lang$^{\rm a}$\thanks{Speaker at the conference.}
and T.~K. Pany\address{Institut f\"ur Theoretische Physik,
Universit\"at Graz, A-8010 Graz, Austria}}
       
\begin{document}

\begin{abstract} 
We determine non-perturbatively a fixed-point (FP) action for fermions
in the two-dimensional U(1) gauge (Schwinger) model.  Our
parameterization for the fermionic action has terms within a $7\times
7$ square on the lattice, using compact link variables. With the Wilson
fermion action as starting point we determine the FP-action by
iterating a block spin transformation (BST) with a blocking factor of 2
in the background of non-compact gauge field configurations sampled
according to the (perfect) Gaussian measure. We simulate the model at
various values of $\beta$ and find excellent improvement for the
studied observables.
\end{abstract}

\maketitle

\section{INTRODUCTION}

We are interested in a lattice representation of the continuum action
for the 2D U(1) gauge theory with massless fermions \cite{Sc62b} (the
Schwinger model). The lattice action should respect the basic
symmetries and should have the correct (naive)  classical limit. Also
one  has to take care for the fermion doubling problem. Beyond these
requirements the form of the lattice action is largely arbitrary.

The traditional lattice actions are ultra-local but have corrections
${\cal O}(a^2)$ (of the lattice spacing constant) for bosons and ${\cal
O}(a)$ for fermions. Improved actions have smaller corrections in
powers ${\cal O}(a^n)$ but introduce more terms. As long as the
contributions are exponentially damped with regard to their extension
in real space one calls the action local. Optimally, an improved action
has no such corrections and thus no corrections to the leading critical
behaviour. Such actions have been called ``perfect''.  More terms
complicate the simulation and one has to find a compromise bet\-ween
efficiency and perfectness.

Various approaches to improvements are reviewed in
\cite{Ni97,Hasenfratz}.  We follow the path inspired by scale
transformations leading to FP-actions \cite{HaNi94}. We determine an
optimal fermion action in the background of gauge field configurations
sampled according to their (optimal) Gaussian measure.  The resulting
FP-action in this approximation defines a classically perfect action
and for large $\beta$ we expect that it is a good approximation for the
renormalized trajectory. In this limit  the gauge field acts like a
background field for the fermionic sector \cite{BiBrCh96}, and the
fermionic action stays quadratic in the  fermionic field variables.
Further details may be found in \cite{PaLa97}. FP-actions for that
model were also studied with the method of small fields
\cite{BiWi96}  and recently in a perturbative
expansion \cite{FaLa97}.

\section{DETERMINATION~OF~THE~FP-ACTION}

We denote the lattice action by
\begin{equation}
\displaystyle\beta\,S(A)-S_F(\bar{\Psi},\Psi,A)\;,\quad
S_F= \bar{\Psi}\,M(A)\,\Psi \;,
\end{equation}
where $S(A)$ denotes the gauge field part, $M$ the lattice Dirac
operator matrix, and $\beta=1/g^2$ is the gauge field coupling.

We block from a so-called fine square lattice with sites $x\in
Z_N\times Z_N$ to a coarse lattice organizing the fine lattice in
$2\times 2$ blocks which constitute the points $x'$ of the coarse
lattice. The Grassmann fields are $\bar{\Psi}(x), \Psi(x)$ 
(respectively $\bar{\Psi}'(x'), \Psi'(x')$ on the coarse lattice); the
real, non-compact gauge fields $A_\mu(x)$ live on the links.  For the
fermions we use anti-periodic boundary conditions and for the gauge
field periodic ones.

The BST is defined as
\begin{equation}
\begin{array}{l}
\label{eq:bst}
\displaystyle
e^{ -\beta'S'(A') + S'(\bar{\Psi}', \Psi', A' )+ c } =
\int D_fA\,D\Psi\, D\bar{\Psi}\\
\quad\times e^{ -\beta\left(S(A)+T(A,A')\right) +
S(\bar{\Psi},\Psi,A)+T(\bar{\Psi},\Psi,\bar{\Psi}',\Psi') }\;.
\end{array}
\end{equation}
We fix the gauge within each block to the so-called {\em fine} gauge,
where the plaquette field strength is distributed
equally among the four link variables \cite{PaLa97}. We integrate  only
over gauge field configurations in the fine gauge, defining the path
integral measure $D_fA$.

The kernel of the BST for the fermions was taken from
\cite{BiWi96} with parameters suitable chosen in order to have
maximum locality in the situation of free fermions.  For the kernel of
the gauge field we define an average over the four 2-link connections
between corresponding sites in adjacent blocks.  More details are
discussed in \cite{PaLa97}. The resulting action on the coarse lattice
is gauge invariant, hermitian invariant, invariant under the charge
conjugation and respects the lattice symmetry. It does violate chiral
symmetry.

There exists a unique minimizing configuration of $S(A)+T(A,A')$ which
we  denote by $A_{\textrm{\scriptsize min}}(A')$.  In our case it can
be computed straightforwardly by solving a set of linear equations. For
$\beta \to \infty$ this saddle point $A_{\textrm{\scriptsize min}}(A')$
dominates the path integral. The Grassmann integration with subsequent
identification of all bilinear terms in the fermionic variables defines
the fermionic block action and the FP equations (in the limit $\beta
\to \infty$).  In order to use the action also at moderate $\beta$
values, we have to calculate the blocked action for strongly
fluctuating configurations $A'$, too. 

\noindent{\bf Gauge field FP-action:}\\
For our BST the ultra-local standard (non-compact) plaquette action is
a FP, up to the wave  function  renormalization \cite{BiWi96}. 
In $d=2$ this action is $S_P(A)=\frac{1}{2}\,\sum_x F(x)^2$.

\noindent{\bf Fermion~FP-action:}\\
We  parameterize
$S'(\bar{\Psi}',\Psi',A')$ with a finite number of coupling constants,
\begin{equation} \label{eq:fermfit}
\begin{array}{l}
S_F(\bar{\Psi},\Psi,A) = \bar{\Psi}\,M_F(A) \,\Psi =\\
\displaystyle\quad
\sum_{i=0}^3\sum_{x\, , f}\,  \rho_i(f)\, 
\bar{\Psi}(x)\, \sigma_i\, U(x,f)\, \Psi(x+\delta f)\;.
\end{array}
\end{equation}
Here $M_F(A)$ is the parameterized fermion matrix, $f$ denotes a closed
loop through $x$ or a path from the lattice site $x$ to $x+\delta f$
(distance vector $\delta f$) and $U(x,f)$ is the parallel transporter along this path. The
$\sigma_i$-matrices denote the Pauli matrices for $i= 1,2,3$ and the
unit matrix for $i=0$.

The conditions $\sum_f \rho_1(f) (\delta f)_1 = 1$ and  $\sum_f
\rho_0(f) = 0$ guarantee, that  $\rho$ is normalized and $S_F$
reproduces the action of the massless model in the naive continuum
limit ( $(\delta f)_1$ denotes the component in the 1-direction). We
first considered terms connecting the central site $x$ with any other
site $x+\delta f$ in a $7\times 7$ lattice.  Invariances  of  the
action under certain  symmetries provide further reductions.  In the
iteration process we found that one may omit several of the
original terms. Altogether we finally considered 33 different geometric
shapes corresponding to 123 independent coupling constants \cite{WWW}.

The FP-action is determined in an iteration procedure,
starting from the Wilson action for the massless Schwinger model
($\kappa  = 1/4$).
\begin{itemize}
\item
We generate 50 gauge field configurations $A'$ on the coarse $7\times
7$ lattice according to their probability distribution $e^{-\beta'
S_P(A')}$. For each of these we then find the minimizing
configuration $A_{\textrm{\scriptsize min}}(A')$.

\begin{figure}[t]
\epsfig{file=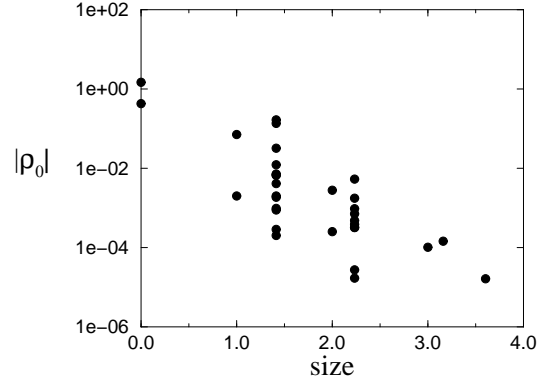,width=7cm}
\vspace{-18pt}
\caption{Logarithmic plot for the absolute values of the couplings
$|\rho_0|$ vs. their lattice extension.}
\label{fig:locality}
\vspace{-9pt}
\end{figure}

\item
With  $A_{\textrm{\scriptsize min}}(A')$ we construct the fermion matrix on
the fine lattice and perform the BST (Grassmann integral)
giving  the $(2\cdot7^2)\times(2\cdot7^2)$ fermion matrix
$M_{BST}(A_{\textrm{\scriptsize min}})$ on the coarse lattice. 
This is done for all
50 gauge field configurations.
\item The resulting fermion matrices are compared with the
fermion matrices $M_F(A')$ for the coarse lattice.
A new set of parameters according
(\ref{eq:fermfit}) is determined by minimizing
\begin{equation}
\sum_{A'} \| M_{BST}\left(A_{\textrm{\scriptsize min}}(A')\right)-M_F(A')\|^2\;,
\end{equation}
for the matrix norm $\|M\|^2\equiv\sum_{i,j}\,|M_{ij}|^2$.
\end{itemize}
These steps are iterated until (typically after 10 iterations) the
couplings remain stable within statistical fluctuations. We worked at
$\beta' = 20$. Part of the observed (small) fluctuations in the
couplings may be due to cancellations of certain terms in the fermionic
action (redundancies). In fig. \ref{fig:locality} we demonstrate the
locality of our FP-action. Our values are comparable or smaller than
those obtained for the free fermion perfect action \cite{BiWi96}.  The
complete set of couplings may be retrieved from \cite{WWW}.

\section{TESTS FOR THE FP-ACTION}

For our check we rely on direct simulations with the FP-action  on
lattices of size $16\times 16$ (for details cf. \cite{PaLa97}). At each
value of $\beta$ considered we generated 10000 gauge field
configurations with appropriate measure and performed the Grassmann
integrals explicitly, i.e. by computing the corresponding determinant
and inverse fermion matrix.  Thus we obtain results for both
situations, the 1-flavour and the 2-flavour model.  For the error
estimates we repeated the procedure several times. We studied
propagators of various ``mesonic'' operators bilinear in the fermion
fields. Comparing the propagators with analogous simulations for
Wilson-fermions we find significantly improved rotational invariance,
demonstrated both in real space and for the energy-momentum dispersion
relations.

\begin{figure}[t]
\begin{center}
\vspace{-9pt}
\epsfig{file=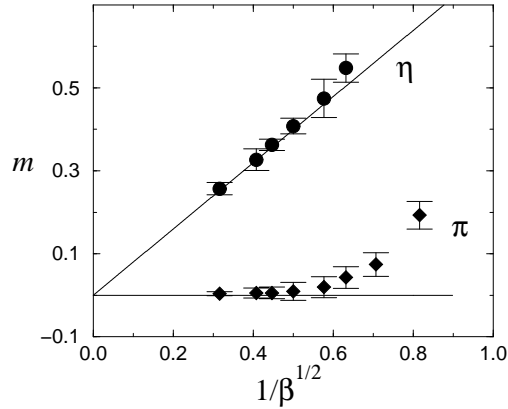,width=6.7cm}
\vspace{-18pt}
\caption{Masses for the massive (full circles) and the massless (diamonds) modes
in the 2-flavour Schwinger model vs. $1/\sqrt{\beta}$. The lines
denote the theoretical expectations for scaling.
\label{Fig_Scaling}}
\end{center}
\vspace{-24pt}
\end{figure}

In the 2-flavour model one expects one massive mode and a massless
flavour-triplet (fig. \ref{Fig_Scaling}). We find non-vanishing
$\pi$-masses only at small $\beta$, indicating deviation of our
FP-action from the renormalized trajectory, i.e. a small signal of
``imperfectness'' to be expected. However, the overall scaling
behaviour predicted from theory (for the 2-flavour model),
$a(\beta)\,m_\eta = \sqrt{{2}/{\pi\,\beta}}$, is nicely recovered for
moderately large values of $\beta>3$.

{\bf Acknowledgement:} We thank W. Bietenholz, F. Farchioni, I.
Hip, U.-J. Wiese and E. Seiler for discussions.

\end{document}